\documentclass[12pt]{article}

\usepackage{a4wide}
\usepackage{cite}
\usepackage{hyperref}
\usepackage{graphicx}
\usepackage{amsmath}
\usepackage{amssymb}
\usepackage{caption}

\def \gsim{\mathrel{\vcenter
     {\hbox{$>$}\nointerlineskip\hbox{$\sim$}}}}
\def \lsim{\mathrel{\vcenter
     {\hbox{$<$}\nointerlineskip\hbox{$\sim$}}}}

\date{\today}

\title{Gluon PDF constraints from the ratio of forward heavy quark
  production at the LHC at $\sqrt{S}=7$ and 13~TeV}

\author{Matteo Cacciari$^{1,2,3,4}$, Michelangelo L. Mangano${^4}$ and Paolo Nason$^5$\\
\footnotesize$^1$Universit\'e Paris Diderot, F-75013, Paris, France\\
\footnotesize$^2$Sorbonne Universit\'es, UPMC Univ Paris 06, UMR 7589, LPTHE, F-75005, 
Paris, France\\
\footnotesize$^3$CNRS, UMR 7589, LPTHE, F-75005, Paris, France\\
\footnotesize$^4$CERN, PH-TH, CH-1211 Geneva 23, Switzerland\\
\footnotesize$^5$INFN, Sezione di Milano Bicocca, Piazza della Scienza 3, 20126 Milan, Italy\\
}

\begin{document}
\maketitle
\vspace{-10cm}
\begin{flushright}
  CERN-PH-TH/2015-171\\
  July 2015
\end{flushright}
\vspace{8cm}

\begin{abstract}
We discuss production of charm and bottom quarks at forward rapidity
in $pp$ collisions at the LHC, updating the QCD predictions for the
run at $\sqrt{S}=13$~TeV. We show that, while the
absolute rates suffer from large theoretical systematics, dominated
by scale uncertainties, the increase relative to the rates 
precisely measured at 7~TeV can be predicted with an accuracy of a few
percent, sufficient to highlight the sensitivity to the
gluon distribution function. 
\end{abstract}

%======================================================================
\section{Introduction}
\begin{figure}[t]
\begin{center}
\includegraphics[width=0.6\textwidth]{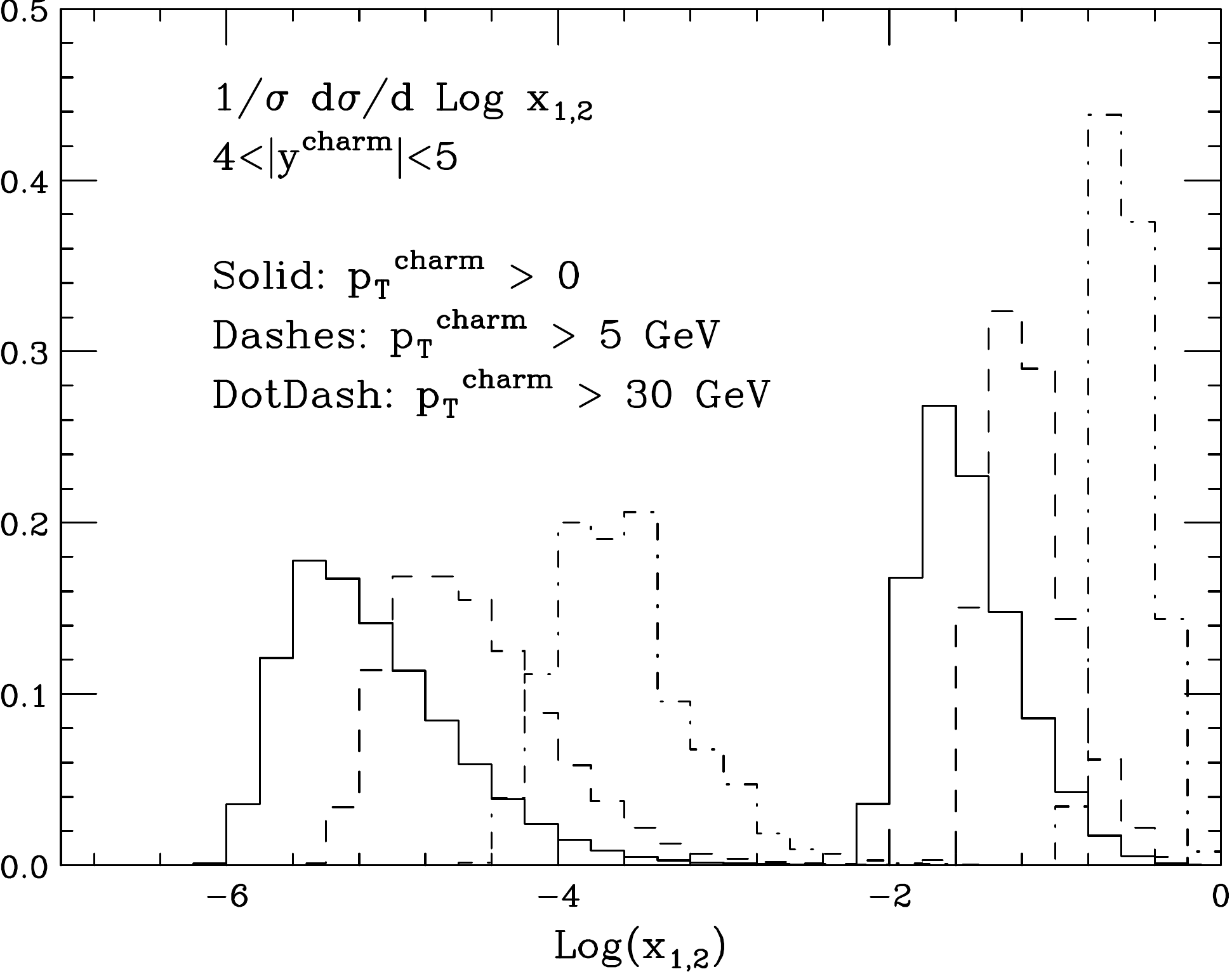}
\caption{\label{fig:x1x2} Distribution of partonic fractional momentum
  $x$ in charm pair production, at LO in $pp$
collisions at $\sqrt{S}=13$~TeV. Events are required to have a charm
quark in the forward region, $4<\vert y \vert <5$. The different
curves correspond to various minimum thresholds for the quark $p_T$,
namely 0, 5 and 30 GeV. All curves are equally normalised.}
\end{center}
\end{figure}

Measurements of heavy quark ($Q=c,b$) production
rates from all LHC experiments during
Run~1\cite{ALICE:2011aa,Abelev:2012pi,Abelev:2012vra,Abelev:2012xe,Abelev:2012gx,Abelev:2012qh,Abelev:2012tca,Aad:2011sp,Aad:2011rr,Aad:2011td,ATLAS:2011ac,ATLAS:2013cia,Aad:2014fpa,ATLAS:2014ala,Khachatryan:2010yr,Khachatryan:2011mk,Khachatryan:2011hf,Chatrchyan:2011pw,Chatrchyan:2011vh,Chatrchyan:2011kc,Chatrchyan:2012dk,Aaij:2010gn,Aaij:2011jh,Aaij:2012jd,Aaij:2012ag,Aaij:2013mga,Aaij:2013noa} have shown
agreement, within the estimated systematics, between data and
theoretical
predictions\cite{Cacciari:2012ny,Kniehl:2011bk,Kniehl:2012ti,Zenaiev:2015rfa}. 
These systematics are typically dominated by theoretical
uncertainties, which are very large: the renormalisation
and factorisation scale dependence, the value of the heavy quark
mass, and, to a lesser extent, the uncertainties of the 
parton distribution functions (PDFs). 
In particular, the scale uncertainty at small transverse 
momentum $p_T$
(namely $p_T\sim m_Q$, where $m_Q$ is the heavy quark mass) reaches values in the range of $\sim 100\%$ in
the case of the charm quark, and of $\sim 50\%$ for the bottom quark.
This situation prevents the use of heavy quarks for precision
measurements. This is frustrating, since the experiments have proven
their ability to measure charm and bottom quarks in regions of small
$p_T$, as well as of large rapidity, where production properties probe
very interesting dynamical regimes, and are sensitive to the gluon
PDFs in both the small- and large-$x$ regions.\footnote{It is a well
  known fact, which we do not need to corroborate further with explicit
  figures, that charm and bottom production in $pp$ collisions is
  fully dominated by $gg$ initial states.} This is shown for
example in Fig.~\ref{fig:x1x2}, where we plot, at leading order (LO) in
QCD, the distribution of
partonic $x$ values in final states of $pp$ collisions at
$\sqrt{S}=13$~TeV where a charm quark is produced in
the rapidity range $4<\vert y \vert < 5$. For small-$p_T$ production,
one probes $x$ values in the region $x\lsim 10^{-5}$, while
for $p_T \gsim 30$~GeV one probes $x\gsim 0.2$. 

In Ref.~\cite{Zenaiev:2015rfa} it was suggested to use, for PDF
studies, only the information encoded in the shape 
of heavy quark differential distributions, since this is more stable
with respect to theoretical uncertainties.
In this work, we show that the predictivity of the theoretical
calculations can be improved by considering appropriate observables,
exploiting the future availability of LHC results at different
energies, namely $\sqrt{S}=13$~TeV in addition to the lower energy
data already available at 7 and 8~TeV.\footnote{In the rest of this
  work we focus on the 7~TeV data, since it is at this energy that
  most -- if not all -- of the charm and bottom production
  measurements have been reported.} The main idea, introduced in a
general context in Ref.~\cite{Mangano:2012mh}, is to consider ratios
of kinematical distributions of heavy quarks -- e.g. inclusive $p_T$
and $y$ spectra -- at different energies.\footnote{While this work 
was in preparation, Ref.~\cite{Gauld:2015yia} has also considered
energy ratios in the context of heavy quark production.}
For a given parton-level
kinematics, the structure of the logarithmic dependence on the
renormalisation and factorisation scales is independent of the beam
energy, since it just depends on the partonic momenta. This allows one,
when studying the scale dependence of the cross-section ratios, to
correlate the scale choice made at the two energies, leading to a major
reduction in the sensitivity to the scale variation.  On the other
hand, the same $(p_T,y)$ kinematics selects initial-state partons with
different values of $x$, since at fixed $y$ we have $x\propto
p_T/\sqrt{S}$. This means that, even though the choice of PDF at the
two energies must be correlated, PDFs having different $x$ dependence
will predict different values for the cross-section ratios, leading to
possibly useful constraints for PDF fits. Other parameters such as the
heavy quark mass, or, when it comes to the complete prediction of
realistic final states, fragmentation fractions to specific hadrons,
fragmentation functions and decay branching ratios, are also fully
correlated at different energies and lead to totally negligible
systematics in the cross section ratios.

The main outcome of these considerations is that, on the basis of the
measurements already performed at 7~TeV, one can predict with much
greater accuracy the cross sections at 13~TeV, and possibly be
sensitive to the (mostly gluon) PDF in regions where it is not yet well
constrained by data. In the rest of the paper we analyse more
quantitatively these statements, focusing on the potential of the LHCb
and, partly, ALICE experiments to combine results from the
forthcoming 13~TeV runs and previous 7 or 8~TeV runs. As part of this
work, we also update to 13~TeV the complete predictions for absolute
cross sections presented in Ref.~\cite{Cacciari:2012ny}. 
We refer to this work for the detailed description of the general
framework of our calculations, and for references to the earlier
literature.

\begin{figure}[t]
\begin{center}
\includegraphics[width=\textwidth]{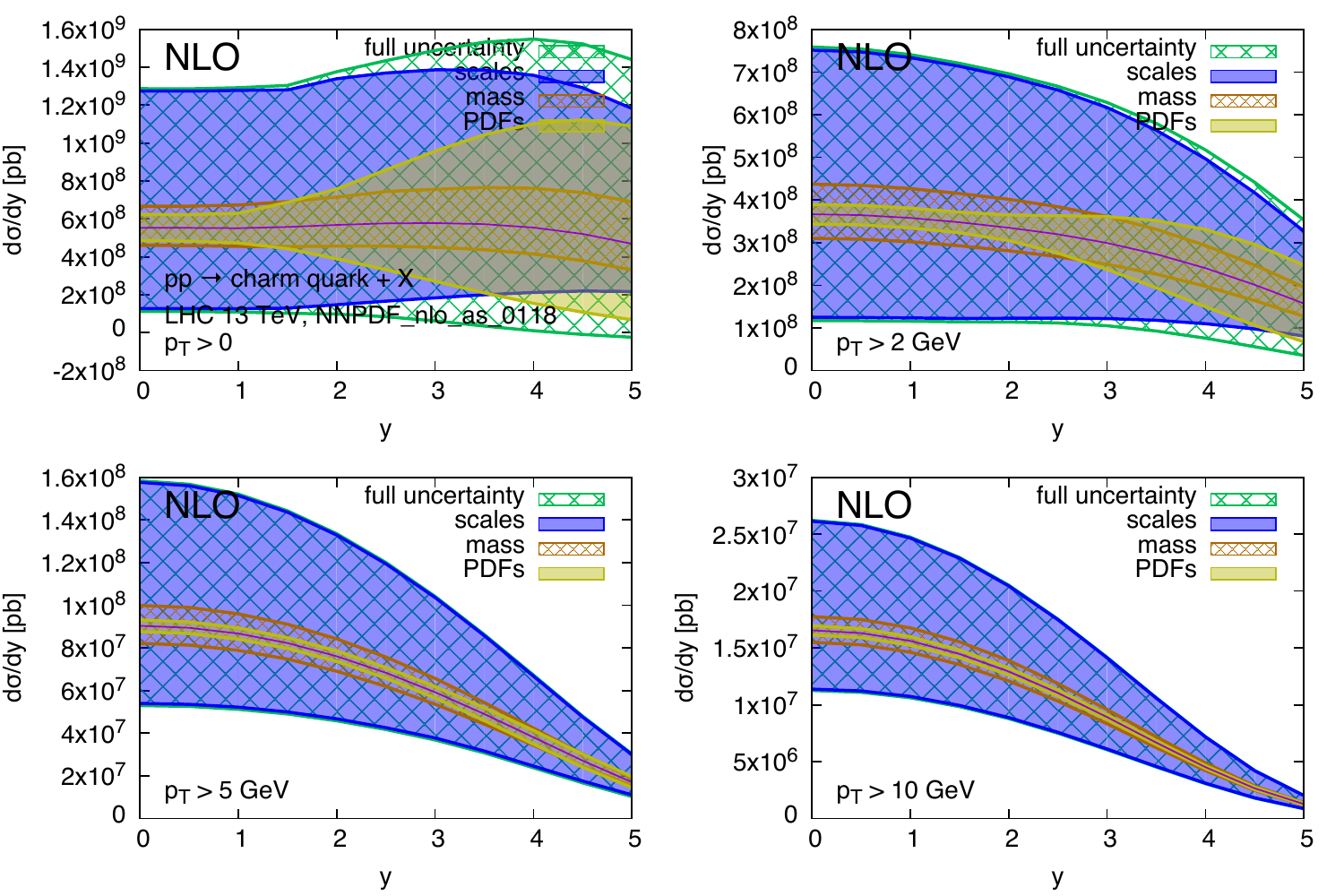}
\caption{\label{fig:charm-xsect} Charm quark rapidity distributions at
  $\sqrt{S}=13$~TeV.}
\end{center}
\end{figure}
\begin{figure}[h]
\begin{center}
\includegraphics[width=\textwidth]{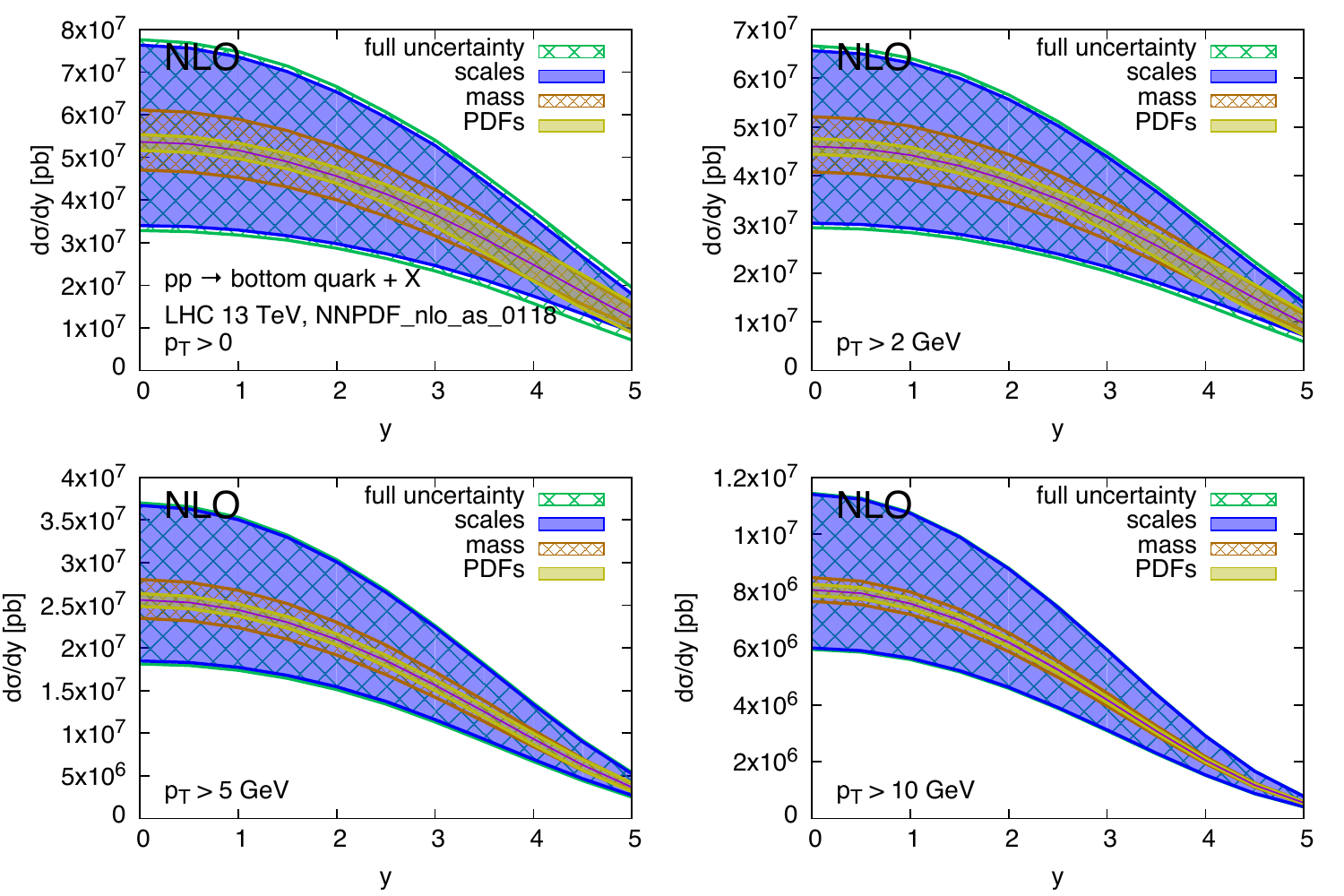}
\caption{\label{fig:bottom-xsect} Bottom quark rapidity distributions
  at $\sqrt{S}=13$~TeV.}
\end{center}
\end{figure}
\begin{figure}[t]
\begin{center}
\includegraphics[width=\textwidth]{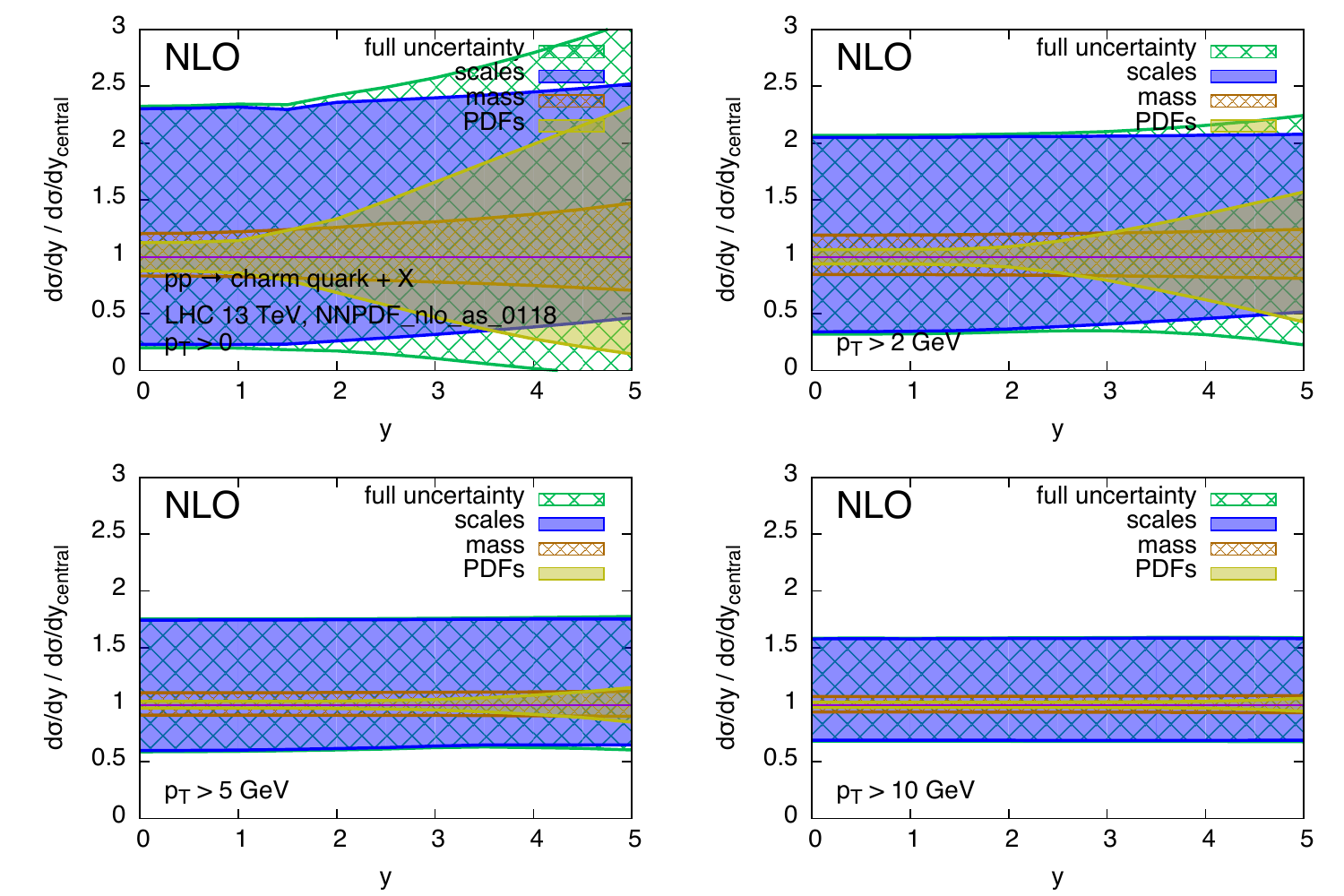}
\caption{\label{fig:charm-band}  Charm quark rapidity distributions at
  $\sqrt{S}=13$~TeV, normalised to the central theoretical prediction.}
\end{center}
\end{figure}
\begin{figure}[h]
\begin{center}
\includegraphics[width=\textwidth]{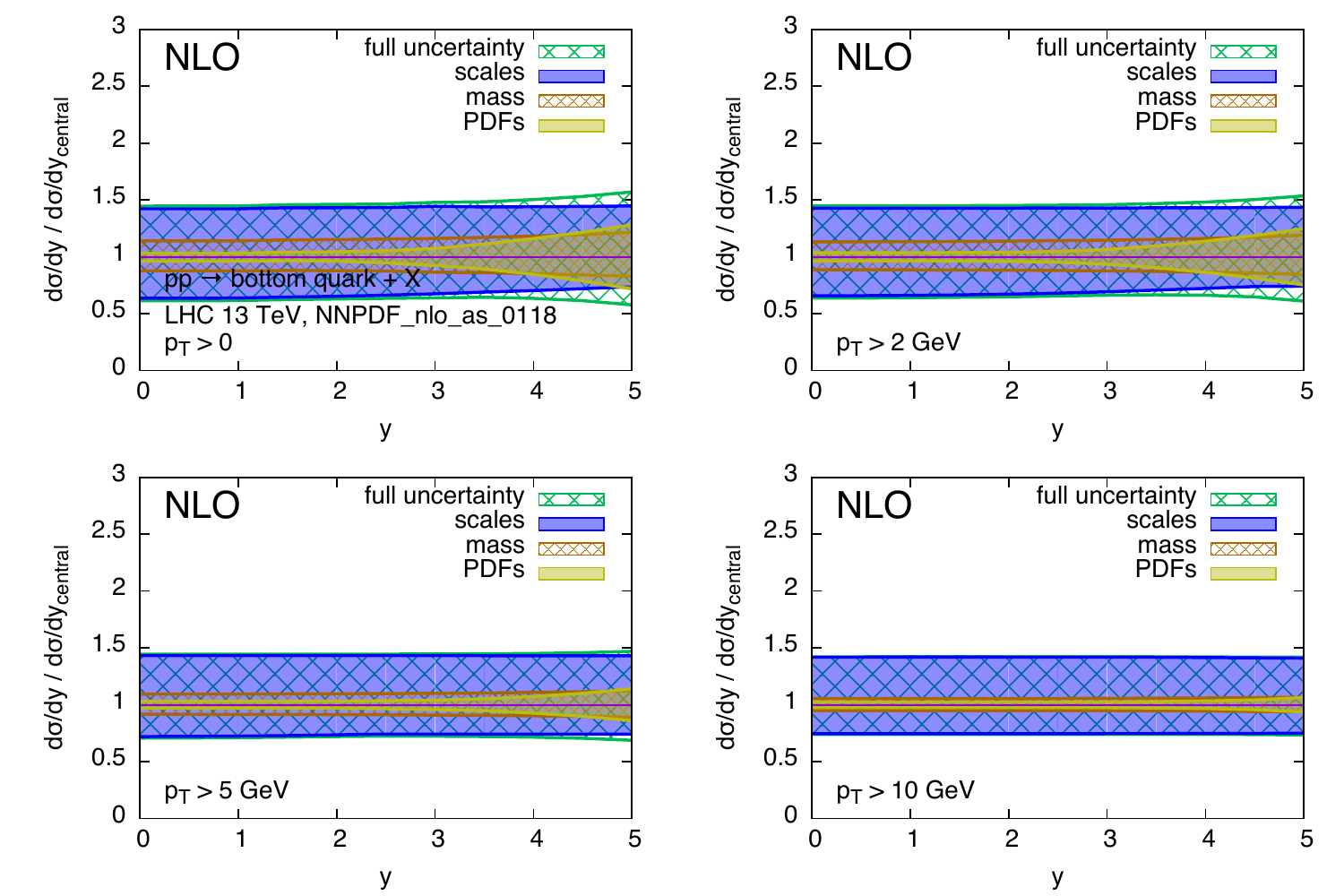}
\caption{\label{fig:bottom-band}  Bottom quark rapidity distributions
  at $\sqrt{S}=13$~TeV, normalised to the central theoretical prediction.}
\end{center}
\end{figure}

\section{General considerations}
The strong scale dependence in charm and bottom pair production is
mostly the consequence of the large
corrections\cite{Nason:1987xz,Nason:1989zy,Beenakker:1988bq}  at the
next-to-leading-order (NLO), and possibly beyond. This is due in part
to the intrinsically large value of $\alpha_S(\mu)$ at the relevant
scales $\mu\sim m_Q$, and in part to the emergence of new processes at
${\cal O}(\alpha_S^3)$.  The large uncertainty can be mitigated in the
regime of $p_T \gg m_Q$, where the dominant higher-order contributions
have a universal logarithmic behaviour that allows for their
resummation\cite{Cacciari:1998it}. 
At lower $p_T$ values, where we can only rely on the fixed-order NLO
QCD calculation,\footnote{NNLO results for heavy-quark pair production
  are in principle known\cite{Czakon:2013goa}. 
In practice, their use is limited today to
  the case of the heavy top quark, due to the intrinsic instability of
  the numerical evaluations for masses in the few-GeV range. The
  extension of the NNLO results to this light mass range will
  therefore require future dedicated numerical studies by the authors
  of the original NNLO calculations.}  the scale dependence reaches
values in the range of $\sim 100\%$ in the case of the charm quark,
and of $\sim 50\%$ for the bottom quark. This situation is shown in
more detail in Figures~\ref{fig:charm-xsect} and
\ref{fig:bottom-xsect}. These show, for $pp$ collisions at $\sqrt{S} =
13$~TeV,\footnote{The situation at a centre-of-mass energy of 7 TeV,
  not shown, is qualitatively identical.} the production cross section
$d\sigma/dy$ for charm and bottom quarks, calculated at the NLO.  The
scale uncertainty is estimated using, as usual,
the envelope of the 7-point scale variation:
\begin{equation}
\label{eq:scalerange}
(\mu_R,\mu_F)=[(1/2,1/2),(1,1/2),(1/2,1),(1,1),(1,2),(2,1)]\times m_T\, ,
\end{equation} 
with $m_T = \sqrt{m_Q^2 + p_T^2}$. 
The scale uncertainty easily dwarves all other sources of uncertainties, namely heavy quark
mass value and PDFs. This is especially true at small transverse
momentum $p_T$ and central rapidity $y$.  Figures \ref{fig:charm-band}
and \ref{fig:bottom-band} show the same data but normalised to the
central theoretical prediction. The relative size of the various
uncertainties can be better appreciated here.

\begin{figure}[t]
\begin{center}
\includegraphics[width=\textwidth]{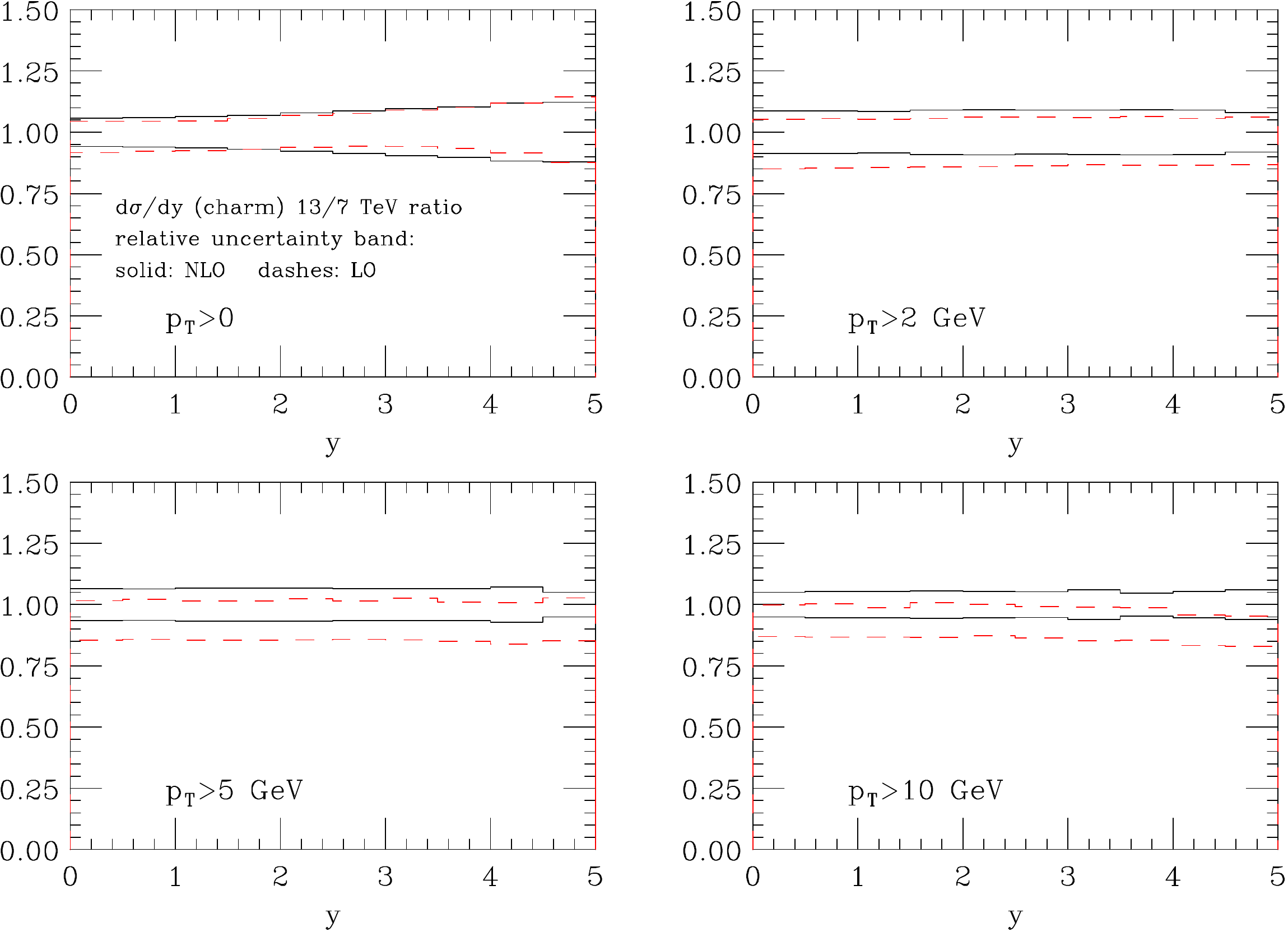}
\caption{\label{fig:lovsnlo} Relative uncertainty in the ratios of
  the charm rapidity differential distribution between 13 and 7 TeV,
  for different $p_T$ ranges. The solid (dashed) lines represent the spread of
  scale dependence at NLO (LO), relative to the centre of the NLO band.}
\end{center}
\end{figure}

As anticipated in the Introduction, we shall consider here the ratio of
differential distributions between different $\sqrt{S}$ values. In
particular, we focus on the rapidity spectra and define:
\begin{equation}
R(y) \equiv
\frac{d\sigma/dy \,(13\,\mathrm{TeV})}{d\sigma/dy \,(7\,\mathrm{TeV})}
\end{equation}
We argued before that, in the evaluation of the scale uncertainty, it
is justified to correlate the scale choices made at the two
energies. This is because the scale of the process is in fact
independent of the collision energy, and it is mainly a function of the
process transverse kinematics.

Higher-order corrections not directly related to the
regularization process could be more or less enhanced by a
higher-energy regime. This is in principle the case of the so-called
small-$x$ logarithmic effects~\cite{Catani:1990eg,Ball:2001pq}. 
However, the evolution in energy from
7 to 13 TeV is not sufficient to expose large effects. 
As a sanity check of this statement, we compare the NLO predictions for
the $R(y)$ ratios to the LO ($R^{\rm LO}(y)$) ones.
Defining $R^{(\rm LO)}_{\rm max,\rm min}(y)$ as the upper and lower extremes in the
envelope of $R^{(\rm LO)}(y)$ values obtained by scanning over the scales choices
given in Eq.~(\ref{eq:scalerange}), and defining the centre of the
NLO envelope by $R_{0}(y)=(R_{\rm max}(y)+R_{\rm min}(y))/2$,
Fig.~\ref{fig:lovsnlo} shows the following distributions, for
different charm $p_T$ thresholds ($p_T>0, \, 2,\, 5, \, 10$~GeV):
\begin{eqnarray}
\mathrm{Solid~lines:} \quad 
\frac{R_{\rm max}(y)}{R_0(y)}  
\quad \mathrm{and} \quad
\frac{R_{\rm min}(y)}{R_0(y)}  \; ,
\nonumber \\
\mathrm{Dahed~lines:} \quad 
\frac{R^{\rm LO}_{\rm max}(y)}{R_0(y)}  
\quad \mathrm{and} \quad
\frac{R^{\rm LO}_{\rm min}(y)}{R_0(y)}  \; .
\nonumber
\end{eqnarray}
Figure~\ref{fig:lovsnlo} shows that
our assumption that a standard scale variation with the \emph{same}
choice of the scales in the numerator and denominator yields a
reasonable estimate of the perturbative uncertainty,
 is verified when going from a LO to an NLO
calculation. In fact, the difference between the LO and NLO result is
always well below the error that is estimated using the scale variation in
the LO result. In other words, $R(y)$ is very stable with respect to
radiative corrections, in spite of the fact that the NLO 
leads to a $K$ factor larger than 2 for the absolute rates. 
We also notice that the NLO uncertainty on $R(y)$ is only mildly reduced
with respect to the LO one, and mostly for the higher $p_T$ values. The
effect is minimal since, as is well known for charm production, the
absolute scale uncertainty at NLO is of the same order of magnitude than at LO.
Nevertheless, we see confirmed the expectation that the scale dependence of
the ratio is much smaller than that of the rates at the individual
energies, being in the 5-10\% range, depending on $p_T$ and $y$. 

\begin{figure}
\includegraphics[width=\textwidth]{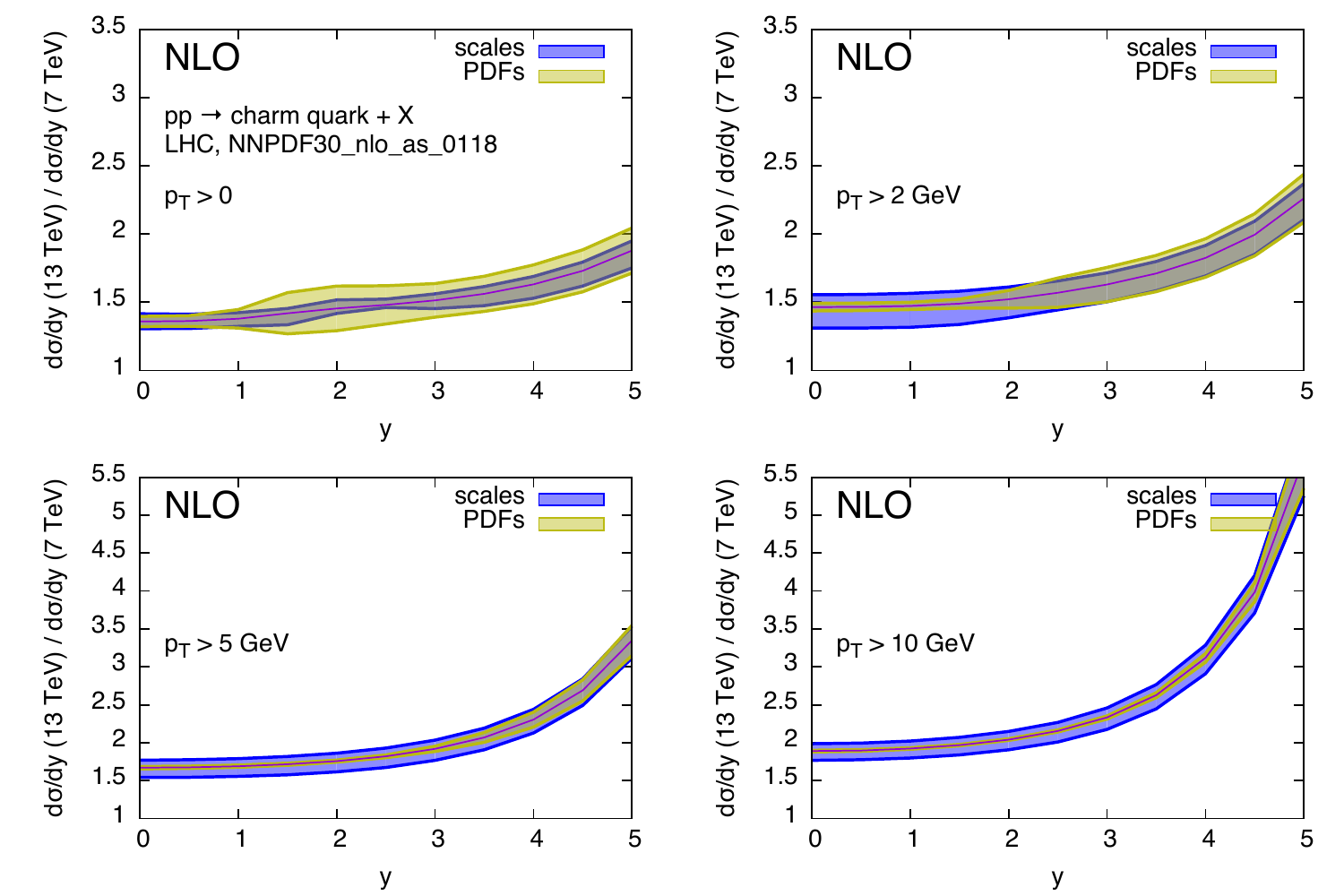}
\caption{\label{fig:charm-ratio}  Ratio of charm quark rapidity distributions 
in $pp$ collisions at $\sqrt{S}=13$~TeV and $\sqrt{S}=7$~TeV collisions in the LHC.}
\end{figure}
\begin{figure}
\includegraphics[width=\textwidth]{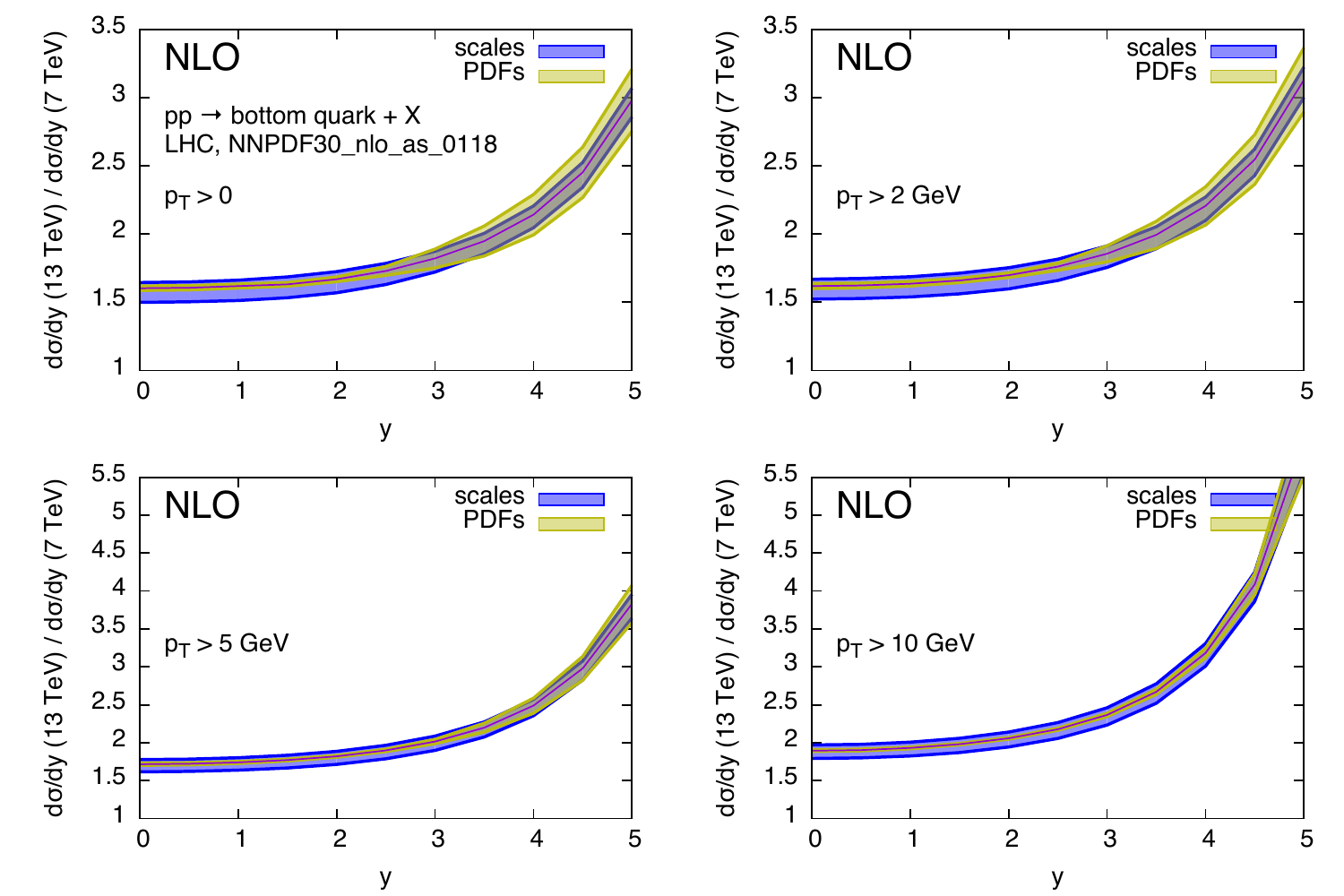}
\caption{\label{fig:bottom-ratio} Ratio of bottom quark rapidity distributions 
in $pp$ collisions at $\sqrt{S}=13$~TeV and $\sqrt{S}=7$~TeV collisions in the LHC.}
\end{figure}

\begin{figure}
\includegraphics[width=\textwidth]{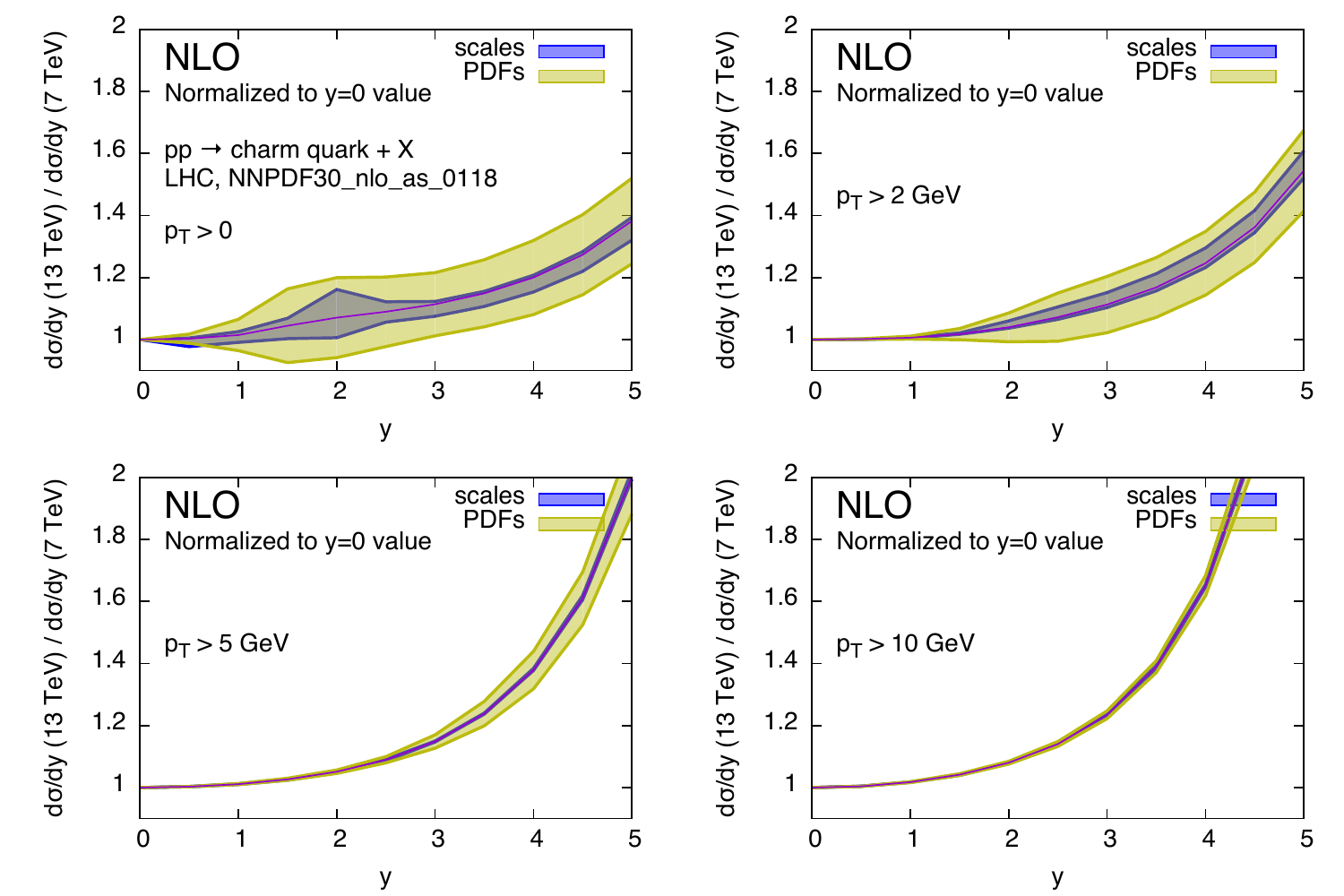}
\caption{\label{fig:charm-ratio-norm} Same as Fig.~\ref{fig:charm-ratio}, but with
further normalisations to the values of the ratios at $y=0$.}
\end{figure}
\begin{figure}
\includegraphics[width=\textwidth]{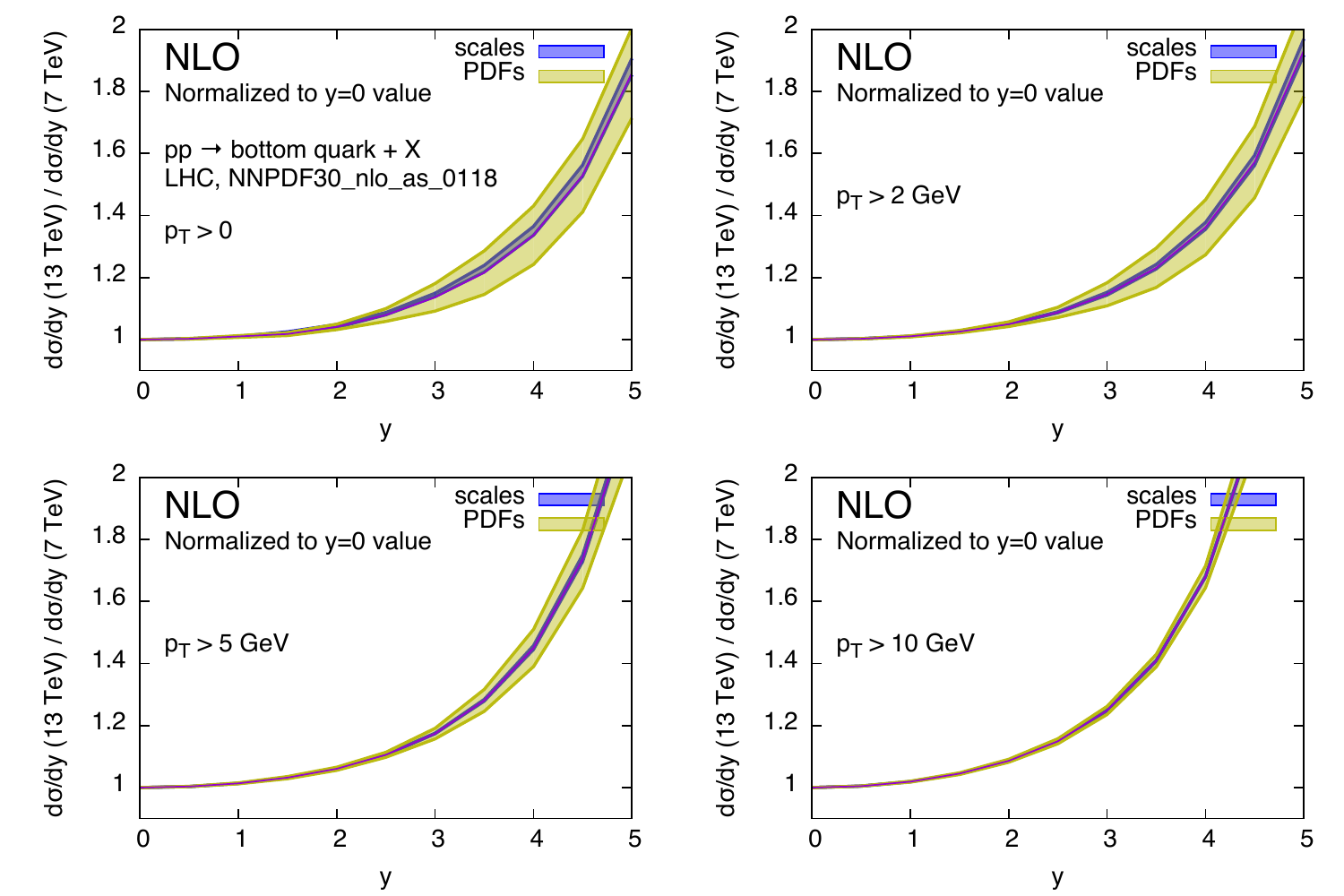}
\caption{\label{fig:bottom-ratio-norm}Same as Fig.~\ref{fig:bottom-ratio}, but with
further normalisations to the values of the ratios at $y=0$.}
\end{figure}

\section{Results}
We show in this Section the complete determination of the ratio
systematics, including the other sources of uncertainty. From now on
we only consider the NLO results for $R(y)$. The ratios
$R(y)$ are calculated using at both centre-of-mass energies the same
renormalisation/factorisation scales, the same mass values and the
same PDF set members. As a reference PDF set for this study, we use the recent
NNPDF30$\_$nlo$\_$as$\_$0118~\cite{Ball:2014uwa}, as implemented in
LHAPDF~\cite{Buckley:2014ana}.

The $R(y)$ distributions, for various $p_T$ thresholds, are shown in
Figures~\ref{fig:charm-ratio}
and \ref{fig:bottom-ratio}. Thanks to the important
suppression of the scale dependence, the overall uncertainties are
greatly reduced, and are now of order 10\% rather than
50-100\%. More importantly, the PDF uncertainty can now become the
dominant one,\footnote{Note that the uncertainty due to the mass value
  is not shown as it is always totally negligible with respect to the
  others.}  if one considers production at sufficiently low
transverse momentum and sufficiently forward rapidity.

In order to reduce uncertainties even further, one can consider taking
a double ratio, and normalise the 13-over-7 cross section ratios to
those measured at a given value of rapidity, as suggested in 
Ref.~\cite{Zenaiev:2015rfa}.
Figures~\ref{fig:charm-ratio-norm} and \ref{fig:bottom-ratio-norm}
show that, when this is done (in this case using $y=0$ as a
reference point) the PDF uncertainty remains by far the dominant one.

This suggests that the double ratio
\begin{equation}
RR(y,\bar{y})=\frac{R(y)}{R(\bar{y})}
\end{equation}
with $\bar y$ conveniently chosen, can represent a powerful handle for
an even more precise determination of the gluon PDFs.

\begin{figure}
\includegraphics[width=0.49\textwidth]{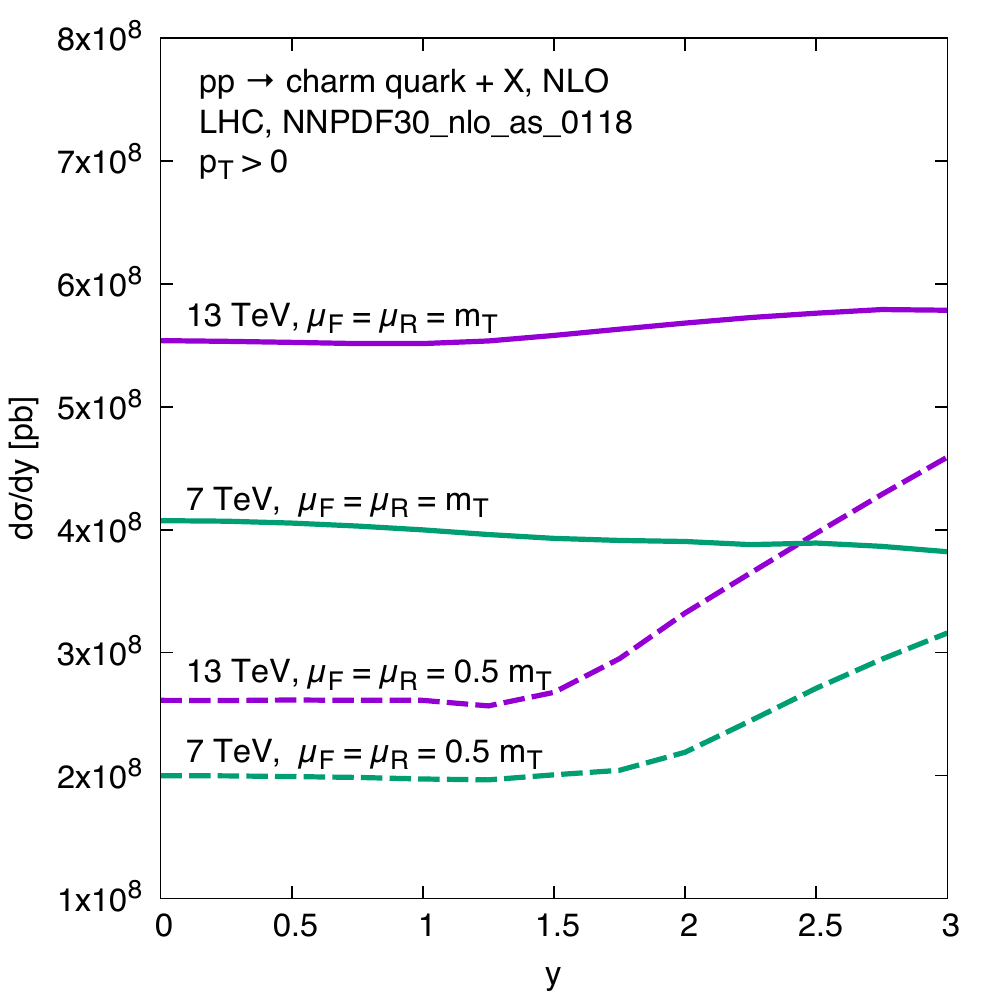}
\includegraphics[width=0.49\textwidth]{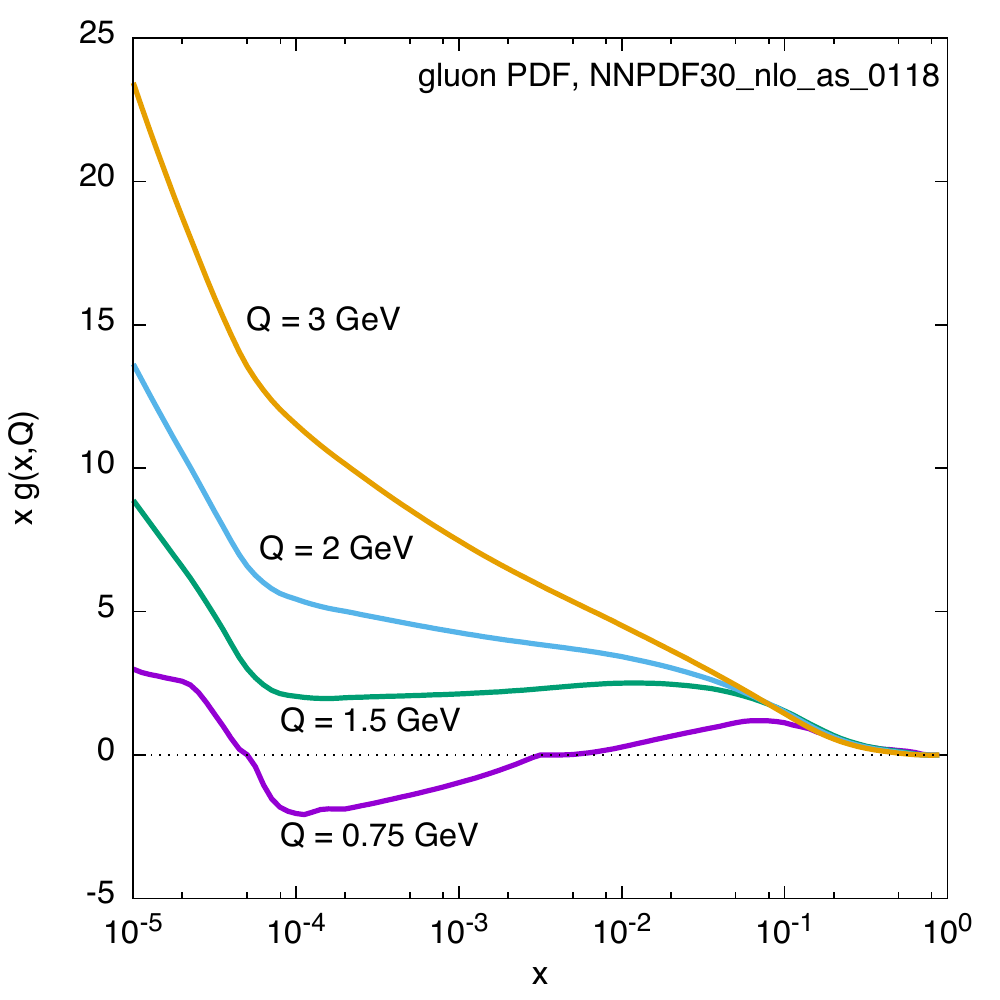}
\caption{\label{fig:small-scale} Left plot: NLO $d\sigma/dy$ distributions for
charm quark production in $pp$ collisions at 7 and 13 TeV centre-of-mass
energies. The curves predicted with a `central' choice, $m_T$, for the factorisation and
the renormalisation scales are shown (solid lines), as well as those given by the choice
$\mu = m_T/2$ (dashed lines). Right plot: the gluon parton
distribution function of NNPDF30 at small scales.}
\end{figure}

A feature that can be observed in Figure~\ref{fig:charm-ratio-norm} (and, to a lesser
extent, also in Figure~\ref{fig:charm-ratio}) deserves an explanation. One can see how
the scale variation band suddenly grows and then decreases around $y\simeq 2$. We have
examined this behaviour in some detail, and found that it can be attributed to the
behaviour of the gluon PDF at very small factorisation scale.
Figure~\ref{fig:small-scale} (left) shows the rapidity distribution of charm
(integrated down to $p_T = 0$) at 7 and 13 TeV centre of mass energy and for two
factorisation and renormalisation scale choices. For the low scale
choice $\mu_F = \mu_R = m_T/2$ the rapidity distribution can be seen to start growing around
$y\simeq 1.5$ and $y\simeq 2$ at 13 and 7 TeV respectively. This behaviour can be understood by looking at the gluon
density displayed in Fig.~\ref{fig:small-scale}~(right): at small scales it displays a very
different slope below and above $x\simeq 10^{-4}$. As the rapidity of the charm quark
increases, the rapidity of the partonic system also increases, thus driving one of two
momentum fractions $x_1$, $x_2$, say $x_1$, towards very small values, while $x_2$
grows. At the larger scale $Q=3\;$GeV this implies that $g(x_1)$ grows and $g(x_2)$
decreases, leading to a roughly constant luminosity, and thus to a constant rapidity
distribution, as can be seen on the left plot. On the contrary, for smaller scales
$g(x_2)$ is flat for $x_2 >10^{-4}$, and the luminosity grows
proportionally to $g(x_1)$, leading to the cross section growth again shown in the
left plot.

As the centre-of-mass energy grows, $x_1$ and $x_2$ are scaled to smaller values, so
that the onset of the effect discussed in the previous paragraph is shifted in
rapidity. This causes a mismatch in the cancellation of scale variation effects
that takes place around $y\simeq 1.5$-2, leading to the feature shown in
Fig.~\ref{fig:charm-ratio-norm}.

It is clear that this effect has to do with the behaviour of the parton
densities at very small scales. On the other hand, we have observed the same
behaviour also with the MMHT PDF set~\cite{Harland-Lang:2014zoa}. In this work
we rely upon the correctness of the PDF's in this region. However, further
investigation of this issue in the framework of PDF fits may be desirable.

\section{FONLL transverse momentum distributions}
The higher centre of mass energy of the Run~2 at the LHC will allow
experiments to measure charm and bottom hadrons up to transverse
momenta much larger than those observed during Run 1. Using an
integrated luminosity of about 15~nb$^{-1}$ at $\sqrt{S}=7$~TeV, LHCb
reported\cite{Aaij:2013mga}
precise spectrum measurements of $D$ mesons up to cross sections
$d\sigma/dydp_T$ of order 1$\mu$b/GeV. This suggests that integrated
luminosities in the range of 1-2fb$^{-1}$
%, as were reached at the end
%of Run~1 and as should be available at Run~2,
should push the
sensitivity up to rates $d\sigma/dydp_T \sim 10$~pb/GeV. This means
$p_T$ values above 30~GeV, even for the largest rapidity ranges
accessible to LHCb. At these
$p_T$ values, much larger than the charm quark
mass, resummation of multiple quasi-collinear gluon emission is
necessary. We therefore provide FONLL\cite{Cacciari:1998it} predictions
for the $p_T$ distributions of charm mesons, and for ratios of $p_T$
distributions, to be compared with
future data. The details of these calculations are documented in
Ref.~\cite{Cacciari:2012ny}. 
We present here the results in graphical form, and collect them in
the Appendix in tabular form for easier use, integrated over finite
bins of $p_T$ and $y$, and inclusive of scale, mass and PDF
uncertainties. 

\begin{figure}
\includegraphics[width=0.5\textwidth]{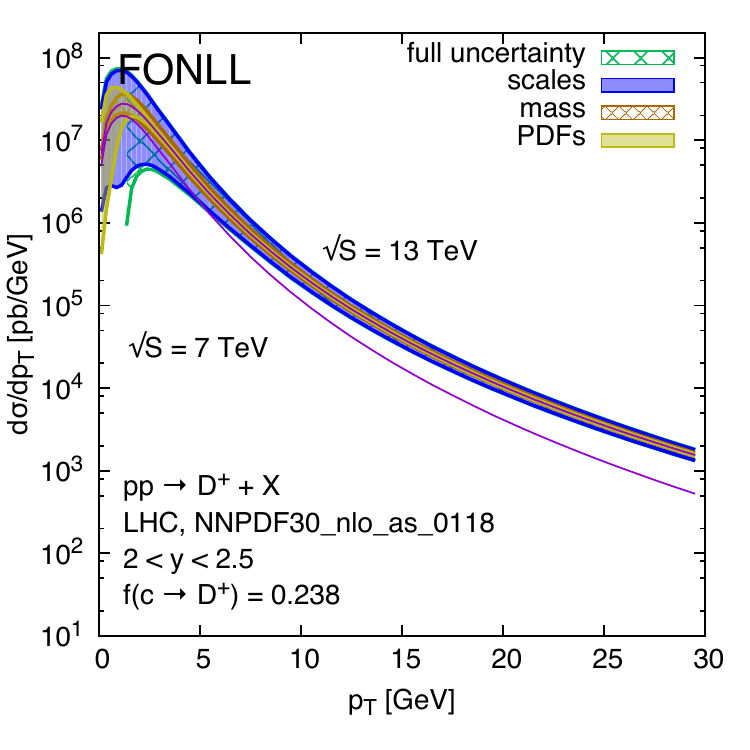}
\includegraphics[width=0.5\textwidth]{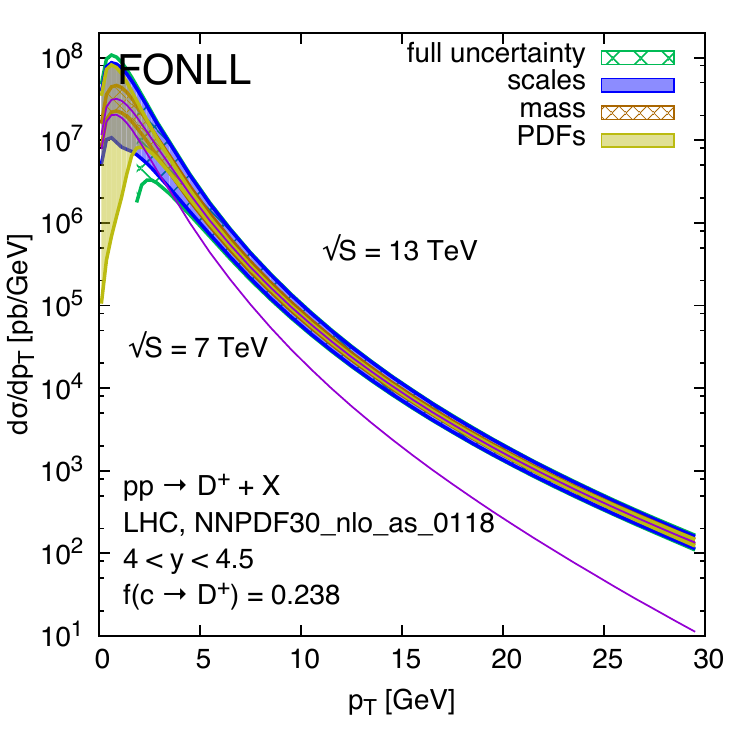}
\caption{\label{fig:D+-ptdist} Transverse momentum distributions of $D^+$ mesons 
in $pp$ collisions at $\sqrt{S}=13$~TeV collisions in the LHC, in the rapidity
regions $2 < y < 2.5$ (left plot) and $4 < y < 4.5$ (right plot). The predictions 
at 7 TeV are also shown for comparison.}
\end{figure}

\begin{figure}
\includegraphics[width=0.5\textwidth]{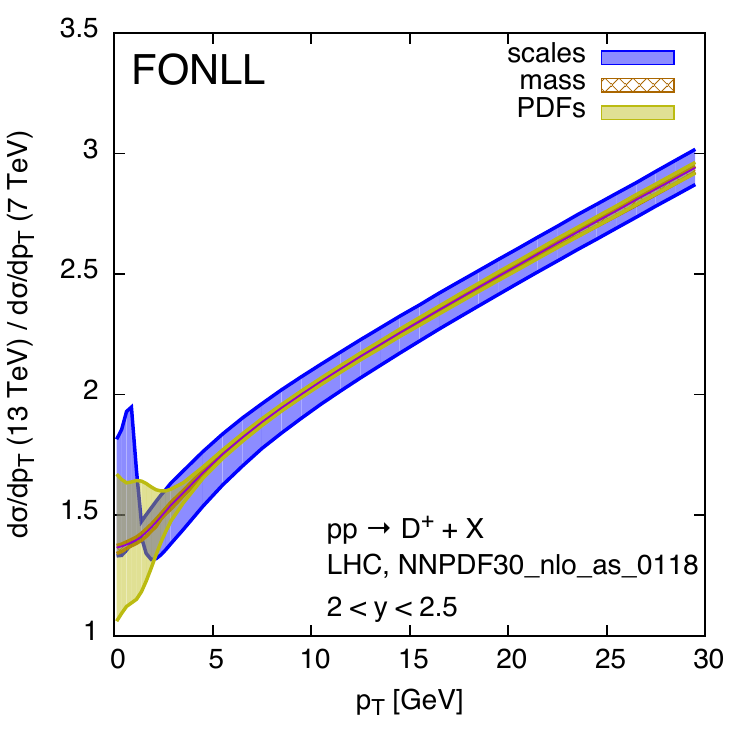}
\includegraphics[width=0.5\textwidth]{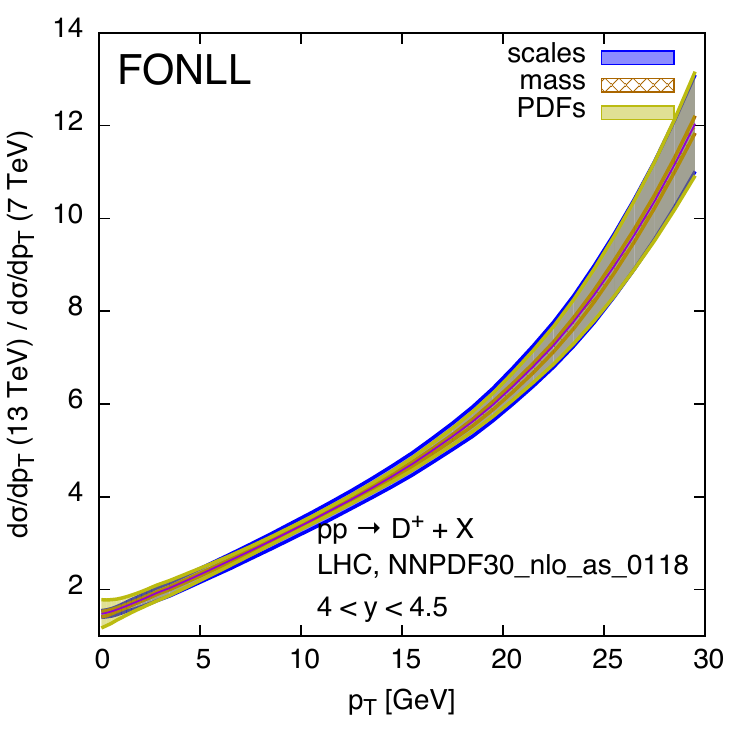}
\caption{\label{fig:D+-ptdist-ratios} Ratios of transverse momentum distributions 
of $D^+$ mesons  in $pp$ collisions at $\sqrt{S}=13$~TeV and $\sqrt{S}=7$~TeV 
collisions in the LHC, in the rapidity regions $2 < y < 2.5$ (left plot) and 
$4 < y < 4.5$ (right plot).}
\end{figure}

Figure~\ref{fig:D+-ptdist} shows the $p_T$ distribution of $D^+$ mesons at $\sqrt{S}=13$ and 7~TeV, in the rapidity ranges $2 < y < 2.5$
and $4 < y < 4.5$, i.e. the first and the last bin where LHCb
reported results during Run~1. 

Figure~\ref{fig:D+-ptdist-ratios} shows the ratio of the same
distributions at 13 and at 7 TeV, showing to what extent the
uncertainties originating from scale variations and from PDFs do
cancel. The interesting observation here is that, in the most forward
rapidity bin, $4 < y < 4.5$, the uncertainty from PDFs is the dominant
one at very small transverse momentum (i.e. $p_T\lsim 5$~GeV), consistently with what we
observed earlier, but it is also still commensurate with the scale
uncertainty at large $p_T$ (i.e. $p_T\gsim 20$~GeV). As shown in Fig.~\ref{fig:x1x2}, in this
high-$y$ and high-$p_T$ region one is in fact probing the gluon
density in the less constrained domain of $x\gsim 0.2$.

Notice that, in order to fully exploit the sensitivity to PDFs via
cross section ratios at high-$p_T$, it will be necessary to increase
the statistics used for the 7 (or 8)~TeV measurements from the ${\cal
  O}(10\, \mathrm{nb}^{-1})$ of the existing publications, to the
${\cal O}(1\, \mathrm{fb}^{-1})$ of the full available dataset.  In
the low-$p_T$ region, on the other hand, it may also be useful to
exploit the double ratios $RR(y,\bar{y})$, as suggested in
Section~3.  A normalisation point at $\bar{y}=0$ can be provided by
measurements from the ALICE experiment, which has sensitivity down to
very low $p_T$ values.

\section{Conclusions}
In spite of the well known limited theoretical precision, we have
given evidence in this paper that measurements of appropriate
observables built upon open heavy quark production rates offer a great
potential for sensitive tests of and constraints on the gluon PDF. The
cross section ratios that we have introduced are very robust with respect
to higher order corrections, in spite of the large $K$ factors that
characterize the absolute rates at a given energy. The step in energy from
$\sqrt{S}=7$ to 13~TeV is large enough that the rate comparisons at
the same value of $y$ and $p_T$ can expose small differences in the
$x$-distributions of the members of current PDF sets. In several
cases, these differences can affect the predictions for cross section
ratios at the 5-10\% level, larger than the other sources of
theoretical systematics. While experimental measurements at this level
of precision are challenging, we expect that suitable analyses of
ratios should be doable, to benefit from similar cancellations of the
systematics of experimental origin.

\section*{Acknowledgements}
M.C. and N.P. are grateful to the Mainz Institute for Theoretical
Physics (MITP) for its hospitality and its partial support during the
completion of this work. 
The work of M.L.M.  is performed in the framework of the ERC grant 291377
``LHCtheory: Theoretical predictions and analyses of LHC physics:
advancing the precision frontier''.

%\clearpage
\section*{Appendix}

\small
This appendix contains the values of the FONLL  predictions for $D^+$ and $B^+$
production cross sections  in $pp$ collisions at 13 TeV, 
and the ratios with the corresponding predictions at 7 TeV. Cross sections and ratios for
$D^0$ and $D^*$ are not included here for brevity, but are qualitatively very similar and can be 
obtained from the authors.

\vspace{1cm}

\tiny

\begin{tabular}{c c c c | c c c c c c c}
\multicolumn{11}{c}{\small FONLL $pp\to D^+ X$,~$\sigma~(13~\mathrm{TeV})$~[pb], $2 < y < 2.5$, $f(c\to D^+) = 0.238$}\\
$p_T^{\rm min}$ & $p_T^{\rm max}$ & $y^{\rm min}$ & $y^{\rm max}$ & central & scales$-$ & scales$+$ & mass$-$ & mass$+$ & PDFs$-$ & PDFs$+$ \\
\hline
0.00 & 0.25 & 2.00 & 2.50 & 1.717$\times 10^6$  & 3.418$\times 10^5$ & 5.914$\times 10^6$ & 1.463$\times 10^6$ & 1.865$\times 10^6$ & 1.047$\times 10^5$ & 4.115$\times 10^6$ \\ 
0.25 & 0.50 & 2.00 & 2.50 & 4.034$\times 10^6$  & 6.671$\times 10^5$ & 1.270$\times 10^7$ & 3.339$\times 10^6$ & 4.608$\times 10^6$ & 3.668$\times 10^5$ & 8.908$\times 10^6$ \\ 
0.50 & 0.75 & 2.00 & 2.50 & 5.522$\times 10^6$  & 7.057$\times 10^5$ & 1.587$\times 10^7$ & 4.453$\times 10^6$ & 6.631$\times 10^6$ & 8.597$\times 10^5$ & 1.081$\times 10^7$ \\ 
0.75 & 1.00 & 2.00 & 2.50 & 6.466$\times 10^6$  & 6.700$\times 10^5$ & 1.735$\times 10^7$ & 5.081$\times 10^6$ & 8.154$\times 10^6$ & 1.953$\times 10^6$ & 1.100$\times 10^7$ \\ 
1.00 & 1.25 & 2.00 & 2.50 & 6.916$\times 10^6$  & 7.147$\times 10^5$ & 1.758$\times 10^7$ & 5.343$\times 10^6$ & 9.011$\times 10^6$ & 3.499$\times 10^6$ & 1.035$\times 10^7$ \\ 
1.25 & 1.50 & 2.00 & 2.50 & 6.869$\times 10^6$  & 8.839$\times 10^5$ & 1.663$\times 10^7$ & 5.276$\times 10^6$ & 9.092$\times 10^6$ & 4.482$\times 10^6$ & 9.258$\times 10^6$ \\
\end{tabular}

\vspace{1cm}

\begin{tabular}{c c c c | c c c c c c c}
\multicolumn{11}{c}{\small FONLL $pp\to D^+ X$,~$\sigma~(13~\mathrm{TeV})$~[pb], $2.5 < y < 3$, $f(c\to D^+) = 0.238$}\\
$p_T^{\rm min}$ & $p_T^{\rm max}$ & $y^{\rm min}$ & $y^{\rm max}$ & central & scales$-$ & scales$+$ & mass$-$ & mass$+$ & PDFs$-$ & PDFs$+$ \\
\hline
0.00 & 0.25 & 2.50 & 3.00 & 2.010$\times 10^6$  & 5.422$\times 10^5$ & 6.881$\times 10^6$ & 1.642$\times 10^6$ & 2.329$\times 10^6$ & 7.547$\times 10^4$ & 5.276$\times 10^6$ \\ 
0.25 & 0.50 & 2.50 & 3.00 & 4.629$\times 10^6$  & 1.069$\times 10^6$ & 1.446$\times 10^7$ & 3.687$\times 10^6$ & 5.591$\times 10^6$ & 2.637$\times 10^5$ & 1.129$\times 10^7$ \\ 
0.50 & 0.75 & 2.50 & 3.00 & 6.157$\times 10^6$  & 1.139$\times 10^6$ & 1.755$\times 10^7$ & 4.810$\times 10^6$ & 7.728$\times 10^6$ & 5.536$\times 10^5$ & 1.345$\times 10^7$ \\ 
0.75 & 1.00 & 2.50 & 3.00 & 6.983$\times 10^6$  & 1.058$\times 10^6$ & 1.859$\times 10^7$ & 5.362$\times 10^6$ & 9.096$\times 10^6$ & 1.136$\times 10^6$ & 1.335$\times 10^7$ \\ 
1.00 & 1.25 & 2.50 & 3.00 & 7.254$\times 10^6$  & 1.053$\times 10^6$ & 1.833$\times 10^7$ & 5.512$\times 10^6$ & 9.684$\times 10^6$ & 2.331$\times 10^6$ & 1.216$\times 10^7$ \\ 
1.25 & 1.50 & 2.50 & 3.00 & 7.028$\times 10^6$  & 1.180$\times 10^6$ & 1.694$\times 10^7$ & 5.334$\times 10^6$ & 9.480$\times 10^6$ & 3.549$\times 10^6$ & 1.052$\times 10^7$ \\
\end{tabular}

\vspace{1cm}

\begin{tabular}{c c c c | c c c c c c c}
\multicolumn{11}{c}{\small FONLL $pp\to D^+ X$,~$\sigma~(13~\mathrm{TeV})$~[pb], $3 < y < 3.5$, $f(c\to D^+) = 0.238$}\\
$p_T^{\rm min}$ & $p_T^{\rm max}$ & $y^{\rm min}$ & $y^{\rm max}$ & central & scales$-$ & scales$+$ & mass$-$ & mass$+$ & PDFs$-$ & PDFs$+$ \\
\hline
0.00 & 0.25 & 3.00 & 3.50 & 2.315$\times 10^6$  & 7.687$\times 10^5$ & 7.868$\times 10^6$ & 1.815$\times 10^6$ & 2.844$\times 10^6$ & 5.524$\times 10^4$ & 6.447$\times 10^6$ \\ 
0.25 & 0.50 & 3.00 & 3.50 & 5.231$\times 10^6$  & 1.522$\times 10^6$ & 1.624$\times 10^7$ & 4.015$\times 10^6$ & 6.669$\times 10^6$ & 1.889$\times 10^5$ & 1.368$\times 10^7$ \\ 
0.50 & 0.75 & 3.00 & 3.50 & 6.773$\times 10^6$  & 1.621$\times 10^6$ & 1.919$\times 10^7$ & 5.131$\times 10^6$ & 8.880$\times 10^6$ & 3.865$\times 10^5$ & 1.605$\times 10^7$ \\ 
0.75 & 1.00 & 3.00 & 3.50 & 7.449$\times 10^6$  & 1.477$\times 10^6$ & 1.972$\times 10^7$ & 5.586$\times 10^6$ & 1.003$\times 10^7$ & 7.226$\times 10^5$ & 1.559$\times 10^7$ \\ 
1.00 & 1.25 & 3.00 & 3.50 & 7.514$\times 10^6$  & 1.399$\times 10^6$ & 1.888$\times 10^7$ & 5.605$\times 10^6$ & 1.028$\times 10^7$ & 1.438$\times 10^6$ & 1.384$\times 10^7$ \\ 
1.25 & 1.50 & 3.00 & 3.50 & 7.095$\times 10^6$  & 1.466$\times 10^6$ & 1.702$\times 10^7$ & 5.310$\times 10^6$ & 9.727$\times 10^6$ & 2.551$\times 10^6$ & 1.164$\times 10^7$ \\
\end{tabular}

\vspace{1cm}

\begin{tabular}{c c c c | c c c c c c c}
\multicolumn{11}{c}{\small FONLL $pp\to D^+ X$,~$\sigma~(13~\mathrm{TeV})$~[pb], $3.5 < y < 4$, $f(c\to D^+) = 0.238$}\\
$p_T^{\rm min}$ & $p_T^{\rm max}$ & $y^{\rm min}$ & $y^{\rm max}$ & central & scales$-$ & scales$+$ & mass$-$ & mass$+$ & PDFs$-$ & PDFs$+$ \\
\hline
0.00 & 0.25 & 3.50 & 4.00 & 2.620$\times 10^6$  & 1.023$\times 10^6$ & 8.851$\times 10^6$ & 1.968$\times 10^6$ & 3.411$\times 10^6$ & 3.891$\times 10^4$ & 7.602$\times 10^6$ \\ 
0.25 & 0.50 & 3.50 & 4.00 & 5.810$\times 10^6$  & 2.027$\times 10^6$ & 1.796$\times 10^7$ & 4.289$\times 10^6$ & 7.816$\times 10^6$ & 1.314$\times 10^5$ & 1.599$\times 10^7$ \\ 
0.50 & 0.75 & 3.50 & 4.00 & 7.326$\times 10^6$  & 2.151$\times 10^6$ & 2.067$\times 10^7$ & 5.372$\times 10^6$ & 1.005$\times 10^7$ & 2.616$\times 10^5$ & 1.847$\times 10^7$ \\ 
0.75 & 1.00 & 3.50 & 4.00 & 7.804$\times 10^6$  & 1.922$\times 10^6$ & 2.056$\times 10^7$ & 5.705$\times 10^6$ & 1.090$\times 10^7$ & 4.662$\times 10^5$ & 1.754$\times 10^7$ \\ 
1.00 & 1.25 & 3.50 & 4.00 & 7.630$\times 10^6$  & 1.747$\times 10^6$ & 1.909$\times 10^7$ & 5.583$\times 10^6$ & 1.072$\times 10^7$ & 8.606$\times 10^5$ & 1.518$\times 10^7$ \\ 
1.25 & 1.50 & 3.50 & 4.00 & 7.009$\times 10^6$  & 1.730$\times 10^6$ & 1.676$\times 10^7$ & 5.167$\times 10^6$ & 9.794$\times 10^6$ & 1.628$\times 10^6$ & 1.244$\times 10^7$ \\
\end{tabular}

\vspace{1cm}

\begin{tabular}{c c c c | c c c c c c c}
\multicolumn{11}{c}{\small FONLL $pp\to D^+ X$,~$\sigma~(13~\mathrm{TeV})$~[pb], $4 < y < 4.5$, $f(c\to D^+) = 0.238$}\\
$p_T^{\rm min}$ & $p_T^{\rm max}$ & $y^{\rm min}$ & $y^{\rm max}$ & central & scales$-$ & scales$+$ & mass$-$ & mass$+$ & PDFs$-$ & PDFs$+$ \\
\hline
0.00 & 0.25 & 4.00 & 4.50 & 2.885$\times 10^6$  & 1.291$\times 10^6$ & 9.729$\times 10^6$ & 2.076$\times 10^6$ & 3.986$\times 10^6$ & 2.637$\times 10^4$ & 8.651$\times 10^6$ \\ 
0.25 & 0.50 & 4.00 & 4.50 & 6.290$\times 10^6$  & 2.551$\times 10^6$ & 1.941$\times 10^7$ & 4.462$\times 10^6$ & 8.944$\times 10^6$ & 8.820$\times 10^4$ & 1.800$\times 10^7$ \\ 
0.50 & 0.75 & 4.00 & 4.50 & 7.721$\times 10^6$  & 2.692$\times 10^6$ & 2.175$\times 10^7$ & 5.474$\times 10^6$ & 1.111$\times 10^7$ & 1.722$\times 10^5$ & 2.045$\times 10^7$ \\ 
0.75 & 1.00 & 4.00 & 4.50 & 7.963$\times 10^6$  & 2.360$\times 10^6$ & 2.092$\times 10^7$ & 5.664$\times 10^6$ & 1.153$\times 10^7$ & 2.987$\times 10^5$ & 1.898$\times 10^7$ \\ 
1.00 & 1.25 & 4.00 & 4.50 & 7.537$\times 10^6$  & 2.068$\times 10^6$ & 1.880$\times 10^7$ & 5.403$\times 10^6$ & 1.086$\times 10^7$ & 5.096$\times 10^5$ & 1.601$\times 10^7$ \\ 
1.25 & 1.50 & 4.00 & 4.50 & 6.724$\times 10^6$  & 1.944$\times 10^6$ & 1.603$\times 10^7$ & 4.874$\times 10^6$ & 9.589$\times 10^6$ & 9.482$\times 10^5$ & 1.277$\times 10^7$ \\
\end{tabular}

\vspace{1cm}

\begin{tabular}{c c c c| c c c c c c c}
\multicolumn{11}{c}{\small FONLL $pp\to D^+ + X$,~$\sigma~(13~\mathrm{TeV})~/~\sigma~(7~\mathrm{TeV})$, $2 < y < 2.5$}\\
$p_T^{\rm min}$ & $p_T^{\rm max}$ & $y^{\rm min}$ & $y^{\rm max}$ & central & scales$-$ & scales$+$ & mass$-$ & mass$+$ & PDFs$-$ & PDFs$+$ \\
\hline
0.00 & 0.25 & 2.00 & 2.50 & 1.3660  & 1.3319 & 1.8150 & 1.3417 & 1.3750 & 1.0613 & 1.6708 \\ 
0.25 & 0.50 & 2.00 & 2.50 & 1.3714  & 1.3341 & 1.8565 & 1.3501 & 1.3809 & 1.0942 & 1.6485 \\ 
0.50 & 0.75 & 2.00 & 2.50 & 1.3785  & 1.3559 & 1.9302 & 1.3630 & 1.3900 & 1.1222 & 1.6348 \\ 
0.75 & 1.00 & 2.00 & 2.50 & 1.3876  & 1.3834 & 1.9488 & 1.3737 & 1.3991 & 1.1369 & 1.6384 \\ 
1.00 & 1.25 & 2.00 & 2.50 & 1.3978  & 1.3967 & 1.6868 & 1.3833 & 1.4102 & 1.1516 & 1.6440 \\ 
1.25 & 1.50 & 2.00 & 2.50 & 1.4119  & 1.4109 & 1.4738 & 1.3957 & 1.4248 & 1.1817 & 1.6421 \\
\end{tabular}

\vspace{1cm}

\begin{tabular}{c c c c| c c c c c c c}
\multicolumn{11}{c}{\small FONLL $pp\to D^+ + X$,~$\sigma~(13~\mathrm{TeV})~/~\sigma~(7~\mathrm{TeV})$, $2.5 < y < 3$}\\
$p_T^{\rm min}$ & $p_T^{\rm max}$ & $y^{\rm min}$ & $y^{\rm max}$ & central & scales$-$ & scales$+$ & mass$-$ & mass$+$ & PDFs$-$ & PDFs$+$ \\
\hline
0.00 & 0.25 & 2.50 & 3.00 & 1.3538  & 1.3369 & 1.4208 & 1.3078 & 1.3816 & 1.0762 & 1.6314 \\ 
0.25 & 0.50 & 2.50 & 3.00 & 1.3678  & 1.3473 & 1.4312 & 1.3264 & 1.3917 & 1.1138 & 1.6218 \\ 
0.50 & 0.75 & 2.50 & 3.00 & 1.3849  & 1.3678 & 1.4406 & 1.3535 & 1.4064 & 1.1612 & 1.6086 \\ 
0.75 & 1.00 & 2.50 & 3.00 & 1.4038  & 1.3957 & 1.4544 & 1.3778 & 1.4250 & 1.2009 & 1.6067 \\ 
1.00 & 1.25 & 2.50 & 3.00 & 1.4250  & 1.4201 & 1.4800 & 1.3997 & 1.4448 & 1.2299 & 1.6200 \\ 
1.25 & 1.50 & 2.50 & 3.00 & 1.4468  & 1.3960 & 1.5141 & 1.4261 & 1.4657 & 1.2542 & 1.6395 \\
\end{tabular}

\vspace{1cm}

\begin{tabular}{c c c c| c c c c c c c}
\multicolumn{11}{c}{\small FONLL $pp\to D^+ + X$,~$\sigma~(13~\mathrm{TeV})~/~\sigma~(7~\mathrm{TeV})$, $3 < y < 3.5$}\\
$p_T^{\rm min}$ & $p_T^{\rm max}$ & $y^{\rm min}$ & $y^{\rm max}$ & central & scales$-$ & scales$+$ & mass$-$ & mass$+$ & PDFs$-$ & PDFs$+$ \\
\hline
0.00 & 0.25 & 3.00 & 3.50 & 1.3693  & 1.3040 & 1.4264 & 1.3123 & 1.4076 & 1.0928 & 1.6458 \\ 
0.25 & 0.50 & 3.00 & 3.50 & 1.3876  & 1.3130 & 1.4363 & 1.3368 & 1.4224 & 1.1333 & 1.6419 \\ 
0.50 & 0.75 & 3.00 & 3.50 & 1.4110  & 1.3290 & 1.4585 & 1.3692 & 1.4412 & 1.1828 & 1.6392 \\ 
0.75 & 1.00 & 3.00 & 3.50 & 1.4378  & 1.3584 & 1.4905 & 1.4011 & 1.4650 & 1.2354 & 1.6401 \\ 
1.00 & 1.25 & 3.00 & 3.50 & 1.4656  & 1.3918 & 1.5268 & 1.4346 & 1.4886 & 1.2818 & 1.6495 \\ 
1.25 & 1.50 & 3.00 & 3.50 & 1.4935  & 1.4070 & 1.5641 & 1.4675 & 1.5156 & 1.3186 & 1.6684 \\
\end{tabular}

\vspace{1cm}

\begin{tabular}{c c c c| c c c c c c c}
\multicolumn{11}{c}{\small FONLL $pp\to D^+ + X$,~$\sigma~(13~\mathrm{TeV})~/~\sigma~(7~\mathrm{TeV})$, $3.5 < y < 4$}\\
$p_T^{\rm min}$ & $p_T^{\rm max}$ & $y^{\rm min}$ & $y^{\rm max}$ & central & scales$-$ & scales$+$ & mass$-$ & mass$+$ & PDFs$-$ & PDFs$+$ \\
\hline
0.00 & 0.25 & 3.50 & 4.00 & 1.4119  & 1.3346 & 1.4761 & 1.3506 & 1.4542 & 1.1266 & 1.6972 \\ 
0.25 & 0.50 & 3.50 & 4.00 & 1.4300  & 1.3439 & 1.4888 & 1.3741 & 1.4698 & 1.1674 & 1.6926 \\ 
0.50 & 0.75 & 3.50 & 4.00 & 1.4574  & 1.3601 & 1.5130 & 1.4075 & 1.4937 & 1.2183 & 1.6965 \\ 
0.75 & 1.00 & 3.50 & 4.00 & 1.4898  & 1.3882 & 1.5489 & 1.4497 & 1.5219 & 1.2753 & 1.7043 \\ 
1.00 & 1.25 & 3.50 & 4.00 & 1.5230  & 1.4241 & 1.5899 & 1.4907 & 1.5516 & 1.3292 & 1.7169 \\ 
1.25 & 1.50 & 3.50 & 4.00 & 1.5574  & 1.4553 & 1.6328 & 1.5269 & 1.5847 & 1.3773 & 1.7375 \\
\end{tabular}

\vspace{1cm}

\begin{tabular}{c c c c| c c c c c c c}
\multicolumn{11}{c}{\small FONLL $pp\to D^+ + X$,~$\sigma~(13~\mathrm{TeV})~/~\sigma~(7~\mathrm{TeV})$, $4 < y < 4.5$}\\
$p_T^{\rm min}$ & $p_T^{\rm max}$ & $y^{\rm min}$ & $y^{\rm max}$ & central & scales$-$ & scales$+$ & mass$-$ & mass$+$ & PDFs$-$ & PDFs$+$ \\
\hline
0.00 & 0.25 & 4.00 & 4.50 & 1.4764  & 1.4013 & 1.5479 & 1.4123 & 1.5257 & 1.1721 & 1.7807 \\ 
0.25 & 0.50 & 4.00 & 4.50 & 1.4975  & 1.4111 & 1.5626 & 1.4349 & 1.5432 & 1.2193 & 1.7756 \\ 
0.50 & 0.75 & 4.00 & 4.50 & 1.5280  & 1.4282 & 1.5893 & 1.4746 & 1.5700 & 1.2749 & 1.7812 \\ 
0.75 & 1.00 & 4.00 & 4.50 & 1.5650  & 1.4569 & 1.6297 & 1.5223 & 1.6038 & 1.3353 & 1.7947 \\ 
1.00 & 1.25 & 4.00 & 4.50 & 1.6052  & 1.4958 & 1.6778 & 1.5655 & 1.6414 & 1.3940 & 1.8164 \\ 
1.25 & 1.50 & 4.00 & 4.50 & 1.6482  & 1.5357 & 1.7287 & 1.6103 & 1.6814 & 1.4506 & 1.8458 \\
\end{tabular}

\vspace{1cm}

\begin{tabular}{c | c c c c c c c}
\multicolumn{8}{c}{\small FONLL $pp\to B^+ X$,~$d\sigma/d\eta~(13~\mathrm{TeV})$~[pb],~$p_T > 0$, $f(b\to B^+) = 0.403$}\\
$\eta$ & central & scales$-$ & scales$+$ & mass$-$ & mass$+$ & PDFs$-$ & PDFs$+$ \\
\hline
0.00 & 1.527e+07  & 9.797e+06 & 2.144e+07 & 1.329e+07 & 1.763e+07 & 1.479e+07 & 1.575e+07 \\ 
0.25 & 1.544e+07  & 9.898e+06 & 2.169e+07 & 1.344e+07 & 1.782e+07 & 1.495e+07 & 1.592e+07 \\ 
0.50 & 1.591e+07  & 1.018e+07 & 2.237e+07 & 1.387e+07 & 1.837e+07 & 1.541e+07 & 1.642e+07 \\ 
0.75 & 1.660e+07  & 1.057e+07 & 2.335e+07 & 1.447e+07 & 1.914e+07 & 1.606e+07 & 1.713e+07 \\ 
1.00 & 1.735e+07  & 1.099e+07 & 2.443e+07 & 1.514e+07 & 1.992e+07 & 1.678e+07 & 1.785e+07 \\ 
1.25 & 1.797e+07  & 1.137e+07 & 2.534e+07 & 1.570e+07 & 2.068e+07 & 1.738e+07 & 1.857e+07 \\
\end{tabular}

\vspace{1cm}

\begin{tabular}{c | c c c c c c c}
\multicolumn{8}{c}{\small FONLL $pp\to B^+
X$,~$d\sigma/d\eta~(13~\mathrm{TeV})~/~d\sigma/d\eta~(7~\mathrm{TeV})$,~$p_T > 0$}\\
$\eta$ & central & scales$-$ & scales$+$ & mass$-$ & mass$+$ & PDFs$-$ & PDFs$+$ \\
\hline
0.00 & 1.6527  & 1.5583 & 1.7069 & 1.6388 & 1.6662 & 1.6353 & 1.6701 \\ 
0.25 & 1.6527  & 1.5574 & 1.7068 & 1.6388 & 1.6655 & 1.6353 & 1.6701 \\ 
0.50 & 1.6509  & 1.5567 & 1.7052 & 1.6372 & 1.6639 & 1.6336 & 1.6683 \\ 
0.75 & 1.6492  & 1.5542 & 1.7094 & 1.6410 & 1.6625 & 1.6308 & 1.6675 \\ 
1.00 & 1.6535  & 1.5521 & 1.7029 & 1.6346 & 1.6671 & 1.6367 & 1.6703 \\ 
1.25 & 1.6490  & 1.5560 & 1.7033 & 1.6344 & 1.6628 & 1.6317 & 1.6664 \\
\end{tabular}

\end{document}